\documentclass{iaus}
\usepackage{graphicx}

\title[Galaxy disks in compact groups] 
{The Evolution of Galaxy Disks\\ in Dense Environments -- \\ Lessons
from Compact Galaxy Groups}

\author[J. Rasmussen et al.]  
{J. Rasmussen$^1$\thanks{Chandra Fellow}, T.J. Ponman$^2$,
L. Verdes-Montenegro$^3$,\\ M.S. Yun$^4$ \and S. Borthakur$^4$}

\affiliation{$^1$Carnegie Observatories, 813 Santa Barbara Street,
Pasadena, CA 91101, USA \\ email: {\tt jr@ociw.edu} \\[\affilskip]
$^2$School of Physics and Astronomy, University of Birmingham,
Edgbaston, \\ Birmingham B15 2TT, UK \\ $^3$Instituto de Astrof\'{\i}sica
de Andaluc\'{\i}a, CSIC, Apdo.\ Correos 3004, E-18080 Granada, Spain\\
$^4$Astronomy Department, University of Massachusetts, Amherst, MA
01003, USA}

\pubyear{2008}
\volume{254}  
\pagerange{000--000}
\jname{The Galaxy Disk in Cosmological Context}
\editors{J. Andersen, J. Bland-Hawthorn \& B. Nordstr{\"o}m, eds.}
\begin{document}

\maketitle

\begin{abstract}
Disk galaxies in compact galaxy groups exhibit a remarkable shortfall
of neutral hydrogen compared to both isolated spirals and spirals in
more loose groups, but the origin of this H{\sc i} deficiency remains
unclear. Based on a sample of highly H{\sc i} deficient compact galaxy
groups, here updated to also include HCG\,58 and 93, we summarise the
first results of a multi-wavelength campaign aimed at understanding
the processes responsible for modifying the H{\sc i} content of galaxy
disks in these environments. While tidal stripping, ram pressure
stripping by hot intragroup gas, and star-formation induced
strangulation could individually be affecting the ISM in some of the
group members, these processes each face specific difficulties in
explaining the inferred deficiency of H{\sc i} for the sample as a
whole. A complete picture of the mechanisms driving the ISM evolution
in the disk galaxies of these groups has thus yet to emerge, but
promising avenues for further progress in this field are briefly
discussed on the basis of the present sample.

\end{abstract}

\vspace{-3mm}

\firstsection 
\section{Introduction}

A substantial fraction of all galaxies in the nearby Universe reside
in a group or cluster environment. Disk galaxies are not only less
common within such environments than in the field, but they also tend
to be deficient in neutral hydrogen. For a group or cluster as a
whole, this H{\sc i}~deficiency $\Delta_{\rm HI}$ can be quantified as
\begin{equation}
  \Delta_{\rm HI} = {\rm log} M_{\rm{HI,pred}} -
  {\rm log} M_{\rm{HI,obs}},
\label{eq,hi_def}
\end{equation}
where $M_{\rm{HI,obs}}$ is the total observed H{\sc i} mass of the
group or cluster, and $M_{\rm{HI,pred}}$ is that predicted for a
corresponding ensemble of isolated galaxies of similar optical
morphology and luminosity.  At the scale of small groups,
comprehensive radio studies have shown such H{\sc i} deficiencies to
be particularly pronounced in \cite{hick82} compact groups (HCGs)
(\cite[Verdes-Montenegro et al.\ 2001]{verd01}). Tidal interactions
can be expected to play a prominent role in affecting the evolution of
galaxy disks in these systems, given the compactness and low velocity
dispersions ($\sigma \sim$ a few hundred km~s$^{-1}$) of the
environment.  However, there is also an indication from {\em ROSAT}
data that significant H{\sc i} deficiency in these groups correlates
with the presence of a detectable X-ray emitting intragroup medium
(IGM), suggesting that galaxy--IGM interactions such as ram pressure
stripping of H{\sc i} may also be important.

Since a non-negligible fraction of all groups in the Universe
(including so-called `fossil' groups) may at some point experience a
phase resembling that of HCGs, an improved understanding of the impact
of the HCG environment on the group members may well have
ramifications extending beyond these somewhat atypical groups. In
order to explore the origin of the shortfall of H{\sc i} in HCGs, and
more generally to shed light on disk galaxy evolution in these
environments, we have therefore embarked on a multi-wavelength
campaign aimed at establishing the detailed properties of the gas and
galaxies in the most H{\sc i} deficient HCGs. The initial focus has
been on the properties of any X-ray emitting IGM, in an attempt to
constrain the importance of galaxy--IGM interactions in these groups.

\section{Sample and Observations}

Our full sample comprises the ten most H{\sc i} deficient systems from
the study of Verdes-\\
Montenegro et al.\ (2001). These all have H{\sc i} masses well below
that expected for their galaxy contents, and so should represent
groups in which the processes destroying H{\sc i} should be in active,
or very recent, operation. {\em Chandra} and {\em XMM-Newton} X-ray
data are now available for eight of these, and this subset is
described in \cite{rasm08}, along with details of the associated X-ray
analysis.  On the radio front, the existing Very~Large~Array data of
the sample have been complemented by single-dish Green Bank Telescope
(GBT) observations, partly to aid in the search for extended, smoothly
distributed H{\sc i} emission within the central group regions
(Borthakur et al., in preparation).  The resulting H{\sc i}
deficiencies range from $\Delta_{\rm HI}=0.27$--1.37, with a
characteristic uncertainty of 0.2, which is dominated by the standard
error on the predicted H{\sc i}~mass. Table~\ref{tab,summary}
summarises the sample and its current X-ray coverage.

\begin{table}
  \begin{center}
  \caption{Summary of group sample and X-ray observations. Distances
  assume $H_0=73$ km s$^{-1}$ Mpc$^{-1}$. Velocity dispersions are
  taken mainly from \cite{ponm96}.}
  \label{tab,summary}
 {\scriptsize
  \begin{tabular}{lccclc}\hline 
  Group & Dist. & $\Delta_{\rm HI}$ & $\sigma$ & X-ray obs. & Expo. time \\ 
                & (Mpc) & & (km s$^{-1}$) & & (ks) \\
  \hline
   HCG\,7     & 54 & 0.60 &  \hspace{0.4mm} 95  & {\em XMM}     & 29 \\
   HCG\,15   & 92 & 0.46 & 404 & {\em XMM}     & 26 \\
   HCG\,30   & 63 & 1.37 &  \hspace{0.4mm} 72  & {\em Chandra} & 29 \\
   HCG\,37   & 97 & 0.33 & 446 & {\em Chandra} & 18 \\
   HCG\,40   & 98 & 0.60 & 157 & {\em Chandra} & 46 \\
   HCG\,44   & 23 & 0.69 & 145 & {\em Chandra} & 20 \\
   HCG\,58   & 89 & 0.51 & 178 & {\em ROSAT}     & 11 \\
   HCG\,93   & 64 & 0.99 & 234 &  {\em ROSAT}/{\em Swift}   & 14/4 \\
   HCG\,97   & 86 & 0.35 & 383 & {\em Chandra} & 36 \\
   HCG\,100  & 69 & 0.27 & 100 & {\em Chandra} & 42 \\ \hline
  \end{tabular}
  }
 \end{center}
\vspace{1mm}
\end{table}

The two groups without {\em Chandra} or {\em XMM} data, HCG\,58 and
93, have been targeted in {\em ROSAT} pointings. As a by-product of
our effort to also obtain UV data for the sample, we have recently
acquired a short {\em Swift} X-ray exposure of HCG\,93, with similar
data underway for HCG\,58. No obvious diffuse X-ray emission is
detected in the {\em Swift} data of HCG\,93, consistent with the
earlier {\em ROSAT} result of \cite{ponm96}, but the sensitivity of
the {\em Swift} data is insufficient to provide strong constraints on
the density of any IGM. The IGM in HCG\,58 also remained undetected in
the {\em ROSAT} analysis of \cite{osmo04}.

\section{Results So Far}

The X-ray analysis of the {\em Chandra}/{\em XMM} subsample
(\cite[Rasmussen et al.\ 2008]{rasm08}) has revealed a remarkable
diversity in hot IGM properties across the sample, with diffuse X-ray
emission in the groups ranging from undetected (e.g., in the highly
H{\sc i} deficient HCG\,30) to similar to that in massive X-ray bright
groups (e.g., in HCG\,97). For HCG\,58 and 93, not included in this
earlier analysis, the existing {\em ROSAT} data can be used to place
constraints on their hot IGM masses.  If assuming IGM temperatures of
$kT= 0.5\pm 0.1$ and $0.6\pm 0.1$~keV as estimated from their galaxy
velocity dispersions (\cite[Osmond \& Ponman 2004]{osmo04}), the {\em
ROSAT} results of \cite{osmo04} and \cite{ponm96} for HCG\,58 and 93,
respectively, translate into $3\sigma$ upper limits to their
0.3--2~keV X-ray luminosity and hot gas mass inside the region of our
GBT coverage ($r\approx 4.5'$) of $L_{\rm X} < 1\times
10^{41}$~erg~s$^{-1}$ and $M_{\rm IGM} < 6\times 10^{10}$~M$_\odot$
(HCG\,58), and
$L_{\rm X} < 2\times 10^{40}$~erg~s$^{-1}$ and $M_{\rm IGM} <
1.5\times 10^{10}$~M$_\odot$ (HCG\,93).
These limits are fairly typical of the X-ray undetected groups in our
sample.

The diversity in hot gas content across the full group sample
immediately suggests that galaxy--IGM interactions may not be dominant
in removing H{\sc i} from the disks of the group members. This
interpretation is supported by Figure~1, which reveals no clear
correlation between measured $\Delta_{\rm HI}$ and characteristic
`mean' ram pressure $\langle P\rangle = \langle n_{\rm IGM} \rangle
\sigma^2$ in each group, where $\langle n_{\rm IGM} \rangle$ is the
inferred mean hot IGM density within the region covered by the GBT
data, and $\sigma$ is the galaxy velocity dispersion.

\begin{figure}
\begin{center}
  \includegraphics[width=90mm]{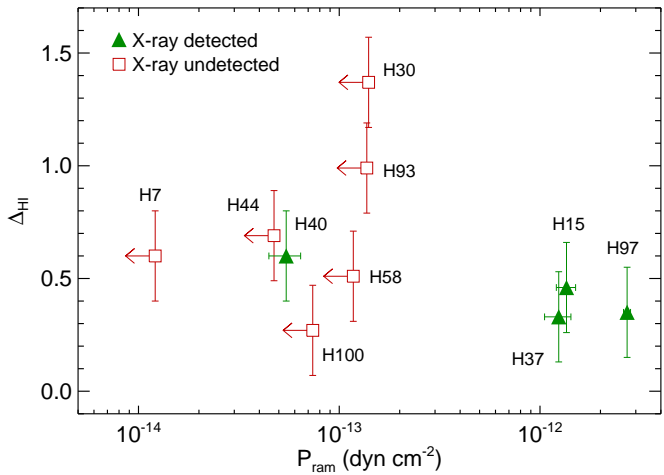}
  \caption{H{\sc i}~deficiencies and characteristic mean ram pressure
  for the various groups. Empty squares represent groups with no
  detectable hot gas.}
  \end{center}
\label{fig,1}
\end{figure}

However, the efficiency of ram pressure stripping depends not only on
the IGM properties but also on the gravitational restoring force of
the individual group members. In order to quantitatively investigate
the importance of stripping in each X-ray detected group, we
constructed a detailed disk--bulge--halo galaxy model, constrained by
the average stellar mass, disk rotational velocity, Hubble type, and
predicted initial H{\sc i} content of our late-type group members (see
Rasmussen et al.\ 2008 for details).  The model was evolved in a
radial orbit within each group gravitational potential as determined
from the X-ray analysis, enabling the IGM ram pressure and H{\sc i}
mass loss due to stripping to be evaluated at each point in the
orbit. Some contribution to H{\sc i} removal could also come from
viscous stripping (Nulsen 1982), which was also included in the model.
Figure~2 compares observed values of $\Delta_{\rm HI}$ to the
resulting model predictions. Uncertainties in model values result from
assuming different initial conditions for the adopted radial
orbits. Observed values of $\Delta_{\rm HI}$ generally exceed modelled
ones, except in HCG\,97 and potentially HCG\,37, suggesting that
galaxy--IGM interactions alone cannot in general explain the observed
H{\sc i} deficiencies, even in the X-ray detected groups.

\begin{figure}
\begin{center}
  \includegraphics[width=70mm]{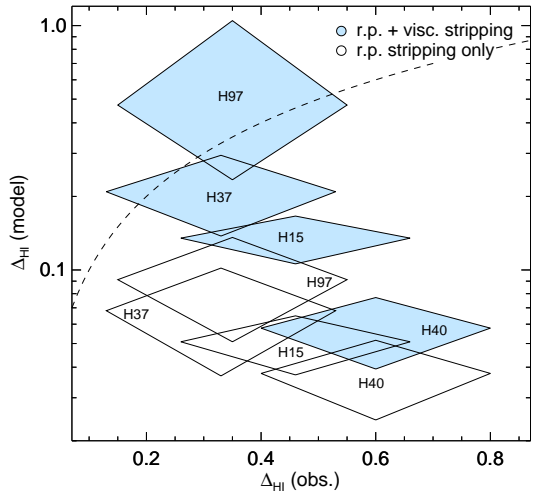}
  \caption{Observed H{\sc i}~deficiencies compared to our stripping
  calculations for the X-ray detected groups. Results are shown for
  ram pressure stripping alone (empty diamonds), and when viscous
  stripping is included (shaded). Dashed line represents equality
  between observed and predicted $\Delta_{{\rm HI}}$.}
  \end{center}
\label{fig,2}
\end{figure}

Such interactions may nevertheless still indirectly affect the H{\sc
i} in spiral disks by facilitating {\em strangulation}. Many disk
galaxy formation models predict that low-redshift spirals, at least
above some threshold mass, are surrounded by hot gaseous halos from
which gas may cool out to provide fuel for ongoing star formation in
the disk (e.g., Toft et al.\ 2002). The removal of this coronal gas by
external forces could contribute to H{\sc i}~deficiency, if the
limited supply of H{\sc i} in the disk is consumed by star formation
without being replenished from the gaseous halo. Detailed
hydro--simulations indicate that this process could be efficient even
in low-mass groups (Kawata \& Mulchaey 2008). Our own stripping
calculations suggest that ram pressure could potentially remove a
sizable fraction of any hot halo gas around the group members in the
X-ray detected groups, from $\sim$30\% in HCG\,40 to $\sim$70\% in
HCG\,97. This process could therefore be important in shutting off the
gas supply that may otherwise ultimately fuel star formation in the
disks of the group members.

\section{Implications and Discussion}

The {\em Chandra} and {\em XMM} data, in most cases representing an
improvement in sensitivity by 1--2 orders of magnitude over the
previous {\em ROSAT} data, manifestly show that the presence of a
substantial hot IGM is not a prerequisite for high H{\sc i} deficiency
in these groups, despite earlier indications of a connection
(Verdes-Montenegro et al.\ 2001).  Although galaxy--IGM interactions
can clearly have affected the H{\sc i} disks of galaxies within X-ray
bright groups such as HCG\,97, such processes cannot generally be
dominant in removing H{\sc i} from the group members in our sample.

Other mechanisms could help account for the observed shortfall of
H{\sc i} within these groups. H{\sc i}~consumption by star formation,
aided by the removal of a continuous supply of H{\sc i} to the disk,
could be playing a prominent role. One problem faced by this scenario,
however, is that the time-scale required to exhaust the H{\sc i}
supply to observed levels (assuming $\Delta_{\rm HI}=0$ initially),
is, on average, at least five~Gyr at the current star formation rates
(SFRs) in the groups. Strangulation may therefore not have been
important in establishing current H{\sc i} levels within our sample.

Another possible explanation for the lack of H{\sc i} are recent tidal
interactions, clearly taking place in some of the groups as evidenced
by Figure~3.  However, analysis of SDSS data has established that such
interactions are commonly accompanied by a clear SFR enhancement
(e.g., Li et al.\ 2008), and yet the SFRs in Hickson groups as
determined from {\em IRAS} fluxes are not elevated relative to values
for field spirals (Verdes-Montenegro et al.\ 1998). Our X-ray analysis
also shows no clear evidence of enhanced nuclear activity within
galaxies in the X-ray brighter or more H{\sc i}~deficient groups,
suggesting that strong nuclear starbursts or AGN activity triggered by
tidally induced gas inflows are not more common or prominent within
the more `evolved' groups. Typical indirect signatures expected of
tidal interactions are thus generally very modest in these groups, and
such interactions may furthermore not themselves destroy H{\sc i}.

In summary, obvious mechanisms that could be invoked to explain the
pronounced deficiency of H{\sc i} observed for the sample clearly each
face some difficulties. The X-ray results indicate that galaxy--IGM
interactions may have played a role in destroying H{\sc i} in the
X-ray bright groups, but it remains unclear whether, for example,
tidal interactions on their own can explain the reduced H{\sc i}
content in the remaining systems.

\section{Prospects for Future Work}

There are several directions in which we hope to take further studies
of these systems in order to shed additional light on the fate of
their missing H{\sc i}. For example, constraints on the typical column
densities of H{\sc i} removed from individual galaxies may allow
estimates of the time-scale for this material to be either
photo-ionized by the intergalactic UV background, or thermally
evaporated in the X-ray bright groups. In the latter case, a detailed
comparison of the X-ray and H{\sc i} morphology of the groups will
also establish to what extent hot and cold intergalactic gas can, in
fact, co-exist in these systems. This should help in building a more
complete picture of the mechanisms affecting any H{\sc i} that is not
being destroyed {\em in situ} within the group members, e.g.\ by star
formation.

Another interesting possibility to explore is related to the evidence
that SFRs in Hickson groups are not globally enhanced relative to the
field. This result could potentially be misleading, perhaps masking an
evolutionary trend in which galaxies joining these groups initially
experience an episode of enhanced star formation which is then
followed by an environment--driven suppression. This possibility could
be reflected in a larger intrinsic variation in specific SFRs compared
to similar field spirals. Since the existing SFR estimates of our
group members are almost exclusively based on {\em IRAS} fluxes, for
which only upper limits are available for many of the galaxies, deeper
complementary data are required to obtain more robust constraints on
SFRs for all group members.  For this purpose, we are currently in the
process of obtaining UV data for all groups in the sample (including
HCG\,58 and 93), using the high-throughput UVOT telescope on {\em
Swift}.  We note that some of our groups have SDSS spectroscopic
coverage, but the SDSS fibres only sample the central region of these
very nearby galaxies, whereas the {\em Swift} data should enable mean
SFR estimates across the full galactic disks.

\begin{figure}
\label{fig,UV}
\begin{center}
  \includegraphics[width=130mm]{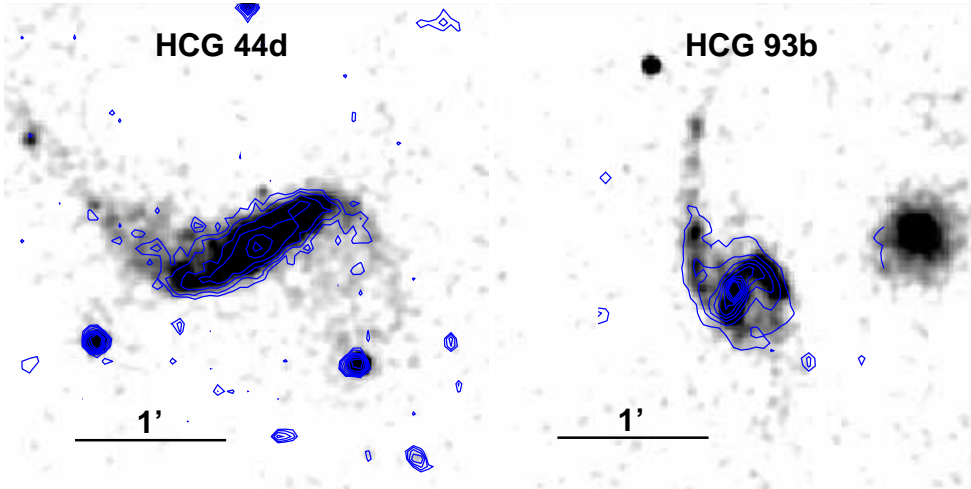}
  \caption{{\em Swift} UV images of HCG\,44d and HCG\,93b with 2MASS
  {\em JHK} contours overlayed. The highly disturbed UV morphologies
  suggest recent tidal interactions in both cases.}.
 \end{center}
\end{figure}

These UV data may also have other useful applications, such as
allowing a search for evidence of recent interactions that have
primarily affected the young stellar component in the disks. As an
example, Figure~3 shows a comparison between the UV and near-infrared
light (dominated by young and old stars, respectively) for two of our
group members which both appear tidally disturbed in optical DSS
images. While the near-infrared light shows a clear warp in both
cases, the UV light appears particularly distorted, suggesting strong
recent interactions in both cases. A comparative lopsidedness analysis
of the UV and NIR light could verify this quantitatively, and should
be possible for most of the bright group spirals given the reasonably
narrow point spread function of UVOT (with FWHM~$\approx 2''$ at UV
wavelengths). Such an approach may aid in quantifying the stage and
time-scale of ongoing tidal interactions, for comparison to the
observed H{\sc i}~deficiencies.

We note in closing that considerable effort has been devoted in the
literature to elucidating the nature and properties of Hickson compact
groups. It is our hope that continued work on the sample discussed
here will ultimately contribute to an improved understanding not only of HCGs
in particular, but of galaxy groups in general, and of the
cosmological evolution of galactic disks in such environments.

\subsection*{Acknowledgements}
Support for this work was provided by the National Aeronautics and
Space Administration through Chandra Postdoctoral Fellowship Award
Number PF7-80050.

\end{document}